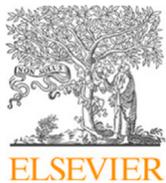

Contents lists available at ScienceDirect

# Electric Power Systems Research

journal homepage: www.elsevier.com/locate/epsr

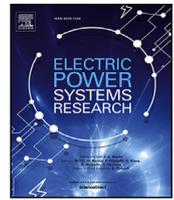

# Evaluation of the power frequency magnetic field generated by three-core armored cables through 3D finite element simulations

Juan Carlos del-Pino-López [a,*], Pedro Cruz-Romero [b], Juan Carlos Bravo-Rodríguez [a]

[a] *Department of Electrical Engineering, Universidad de Sevilla, Virgen de África 7, Sevilla, 41011, Spain*
[b] *Department of Electrical Engineering, Universidad de Sevilla, Camino de los Descubrimientos s/n, 41092, Sevilla, Spain*

## ARTICLE INFO

## ABSTRACT

The great expansion in offshore power plants is raising the concern regarding the cumulative effect of the electromagnetic field emissions caused by submarine power cables. In this sense, owners are required to predict these emissions during the permitting and consenting process of new power plants. This is a challenging task, especially in the case of HVAC three-core armored cables due to their complex geometry. Customarily, 2D approaches based on the finite element method (FEM) have been employed for evaluating the magnetic field emissions caused by these cables. However, inaccurate results are obtained since the phase conductors and armor twisting is omitted. This work develops, for the first time in the literature, an in-depth analysis of the magnetic field caused by this type of cable through an ultra-shortened 3D-FEM model, which is also faced to experimental measurements taken on an actual 132 kV, 800 mm² three-core armored cable. Relevant conclusions are derived regarding the impact of the cable design on the magnetic field emissions, including material properties, as well as single and double-layer armors, presenting the proposed model not only as a valuable tool for predicting purposes, but also for optimizing cable design in terms of magnetic field emissions.

## 1. Introduction

The expansion of offshore wind power plants (OWPPs) worldwide has significantly increased in the last decades, and is expected to continue in the coming years [1,2]. The cumulative effect derived from the increasing number of submarine power cables (SPC) is giving rise to new scenarios with higher electromagnetic field (EMF) emissions over a wider area [3]. This situation causes disturbances to marine life and habitats, although there is a poor understanding of the long-term effects that EMF emissions have on the marine environment, specifically those created by SPC [4–6].

Presently, HVAC cables are the most extended EMF sources in marine applications [2], although the main concern is regarding the magnetic field (MF) emissions [7,8], since the electric field can be shielded by grounding techniques. In this sense, it is a challenging process to calculate the MF emitted by HVAC cables [8–10], especially in the case of three-core armored cables (TCACs) due to their complex geometry, the material properties and, consequently, the electromagnetic interactions that take place inside them (induced sheath currents, flux shunting into the armor wires and eddy currents), where the relative twisting of power cores and armor wires has a key role. These complex interactions have been customarily tackled through 2D numerical simulations, mainly based on the finite element method

(FEM) [11–14]. However, 2D approaches have inherent limitations that strongly influence the results. First, they assume the power cores (and the armor wires) as laid in parallel, although it is well-known that this configuration leads to higher MF emissions than the twisted case [15,16]. Second, they omit the longitudinal component of the MF, so the mitigation effect caused by the armor twisting is not properly evaluated.

Consequently, 3D geometries are required for a proper evaluation of the MF emitted by TCACs. Analytically there have been proposals, but without considering the effect of sheaths and armor [17]. Regarding numerical methods, 3D-FEM simulations are customarily high demanding due to the need of large computational resources [18–21]. This worsens especially when trying to evaluate the MF emissions far from the cable, since this strongly increases the size of the geometry to be simulated. In this sense, the ultra-shortened 3D-FEM model (USM) presented in [22] has drastically reduced these resources, with an overall simulation time of about 1 min. This USM has been faced with experimental measurements at power and harmonic frequencies, providing accurate results regarding the cable sequence impedances, the induced sheath current and total losses [23,24].

Thus, in this study, two real TCACs are considered for developing, for the first time in the literature, an in-depth assessment on the performance of 3D-FEM simulations (particularly the USM) for predicting the impact of TCACs on the marine environment in terms of MF emissions, highlighting the main differences with 2D-FEM results. To this aim, the performance of the USM is extensively assessed through a number of MF measurements obtained in a laboratory setup under different operating conditions. Then, an in-depth parametric analysis is developed to show the impact of the cable design on the resulting MF emissions around TCACs, including aspects never analyzed before through 3D-FEM simulations, such as the relative twisting between armor wires and power cores, as well as the armor layout and its material properties (e.g., single or double armoring, twisting direction, steel wires combined with polyethylene (PE) separators, etc.). Eventually, an application example is developed to show how the USM is a valuable tool for developing new techniques that predict MF emissions, burial depth or SPC detection/tracking [25–27].

\* Corresponding author.
*E-mail address:* vaisat@us.es (J.C. del-Pino-López).

## 2. Evaluating MF emissions through the USM

Due to the symmetries found in both the geometry and the electromagnetic field in a cable length equal to the "crossing pitch" ($CP$) (distance where a phase conductor meets the same armor wire twice), the length of the cable to be employed in 3D-FEM simulations ($L$) can be as short as [22]

$$L = \frac{CP}{N} = \frac{1}{N \cdot \left| \frac{1}{L_c} - \frac{1}{L_a} \right|},$$  (1)

where $N$ is the number of armor wires, and $L_c$ and $L_a$ are the lay length of the phases and the armor wires, respectively (being $L_a < 0$ if twisted in different direction (contralay) and $L_a > 0$ if twisted in the same direction (unilay)). This is achieved by means of boundary conditions with rotated periodicity, that match the magnetic vector potential ($\vec{A}$) at every point ($x^s, y^s, z^s$) belonging to the source boundary with its equivalent ($x^d, y^d, z^d$) at the destination boundary (Fig. 1) when solving the electromagnetic problem

$$\nabla \times \left( \frac{1}{\mu} \nabla \times \vec{A} \right) + j\omega\sigma\vec{A} = \vec{J}_e,$$  (2)

$$\vec{A}\left(x^d, y^d, z^d\right) = \vec{A}\left(x^s, y^s, z^s\right),$$  (3)

being $\omega$ the angular frequency, $\vec{J}_e$ the external current density, and $\mu$ and $\sigma$ the relative permeability and the electrical conductivity of the material, respectively. Due to the rotational twisting of the geometry, a linear coordinate transformation is applied to match the source $\left(\vec{e}_x^s, \vec{e}_y^s, \vec{e}_z^s\right)$ and destination $\left(\vec{e}_x^d, \vec{e}_y^d, \vec{e}_z^d\right)$ coordinate systems through a certain rotating angle $\theta$ (Fig. 1), defined as

$$\theta = 2\pi \frac{L}{L_c}.$$  (4)

On the other hand, to avoid the use of large domains when computing the MF values at a few meters from the TCAC, a non-linear coordinate transformation is applied to the surrounding medium most external layer ("infinite element domain" in Fig. 2), having the effect of stretching it to almost infinity [28]. This way, the impact on the computation burden is very limited.

This USM (implemented in COMSOL Multiphysics [28]) can be solved in less than 1 min [22] in most of the cases, limiting the computational burden when simulating other complex geometries never analyzed before through 3D-FEM simulations, such as those combining steel wires and PE separators or double-layered armored cables. All these cases, together with new features that extensively evaluate the environmental impact of TCACs in terms of MF emissions, are now included in the graphical user interface (GUI) developed recently in [29] (called Virtual Lab). This tool, which is capable of reproducing typical experimental setups usually employed for testing TCACs (with a consequent reduction of cost), serves as a perfect platform for the analysis and optimization of the cable design in terms of power losses, electrical parameters and MF emissions.

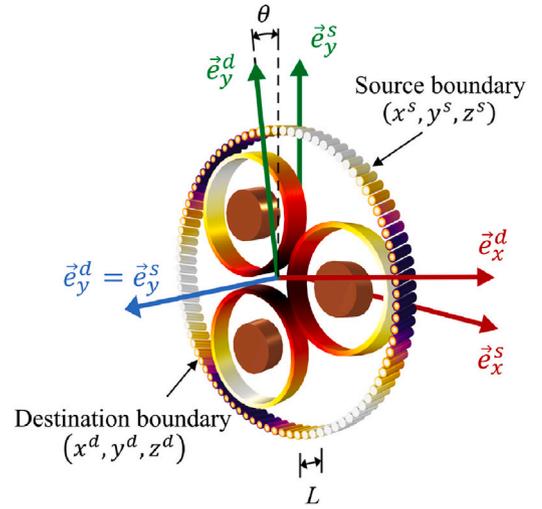

**Fig. 1.** USM: Boundary conditions for applying rotated periodicity in TCACs.

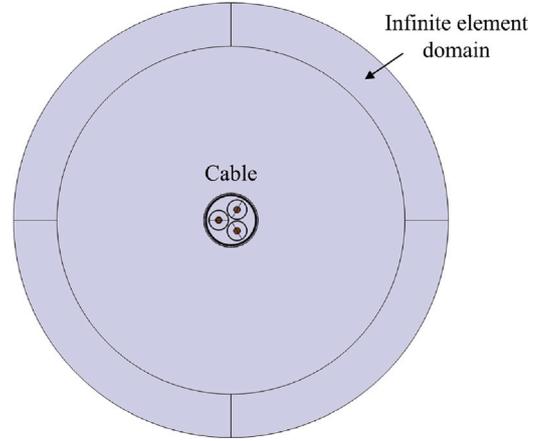

**Fig. 2.** Simulation domains.

### 2.1. Comparison with 2D-FEM simulations

There are two important differences in how 2D-FEM and 3D-FEM models evaluate the electromagnetic interactions inside TCACs. On one hand, 2D geometries assume that the phase conductors, sheaths and armor wires are laid in parallel, leading to higher MF emissions than the twisted configuration, especially at far distance from the cable (solid lines in Fig. 3, obtained for the three conductors of a 132 kV, 800 mm² TCAC through the analytical formulation proposed in [16], which omit the effect of sheaths and armor).

On the other hand, the longitudinal component of the MF is omitted in 2D-FEM models, so that the MF flux lines are confined to a plane perpendicular to the cable axis (Fig. 4a, represented in 3D for untwisted armor wires and power cores for better visualization). Conversely, if twisting is considered (Fig. 4b) the flux lines follow a helical path, flowing longitudinally mainly through the armor wires. This enhances the mitigation effect caused by the flux shunting mechanism, giving rise also to higher eddy currents inside the armor wires (although this has a limited impact on the resulting MF emissions). As a result, a higher mitigation effect is obtained for the 3D (twisted) case than for the 2D approach, as shown in Fig. 3, where the MF emission derived from the USM and 2D-FEM simulations for a 132 kV, 800 mm² TCAC are represented (dashed lines, where the effect of sheaths and armor is now included). As can be seen, greater differences are observed between the black and red dashed lines, mostly at a certain distance from the cable.

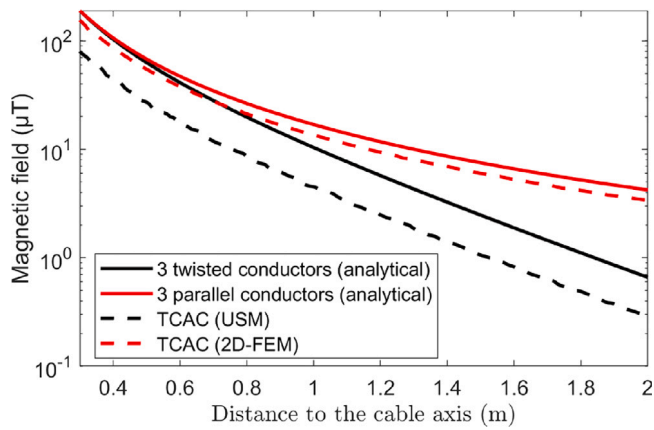

**Fig. 3.** MF obtained with the formulation of [16] for three 800 mm² twisted and parallel conductors, and with the USM and 2D-FEM simulations for a 132 kV, 800 mm² TCAC (phase current of 745 A).

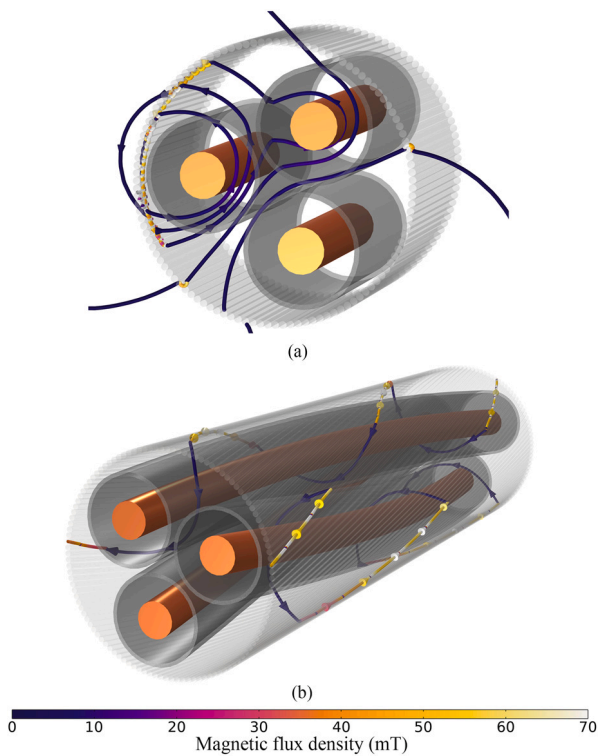

(a)

(b)

**Fig. 4.** Magnetic flux lines in (a) untwisted and (b) twisted TCACs.

These results highlight the importance of a powerful 3D analysis for a proper evaluation of the MF levels around TCACs.

## 3. Case studies

Two HV real TCACs are employed in this work with the two-fold aim of validating the USM through MF experimental measurements and developing an in-depth parametric analysis to show those design parameters that most influence the MF emitted by TCACs. Table 1 summarizes their main dimensions and properties, being $V_n$ the rated voltage, $I_{max}$ the rated current, $S_n$ the cross-section, $D_c$ the conductor diameter, $D_s$, $D_{core}$ and $D_a$ the outer diameter of sheaths, power cores and armor, respectively, $t_s$ the sheath thickness and $d_a$ the armor wire diameter. Both cables have copper conductors and are lead sheathed. Sheaths in solid-bonding (SB) are usually preferred in offshore wind farms, but single-point (SP) configuration is also considered in some studies. Table 1 also includes the material properties considered for each layer, being $\sigma_c$, $\sigma_s$ and $\sigma_a$ the electrical conductivity for the conductors, sheaths and armor wires, respectively. In addition, a non-linear complex permeability ($\mu_r = \mu_r' - j\mu_r''$) is considered for Cable 1 (Fig. 5), corresponding to that of a low-grade steel (LG) [30], while a real value of 300 is considered in Cable 2 for simplicity. Eventually, as in laboratory conditions, it is assumed that tests are performed during short periods, so the cable temperature is mostly uniform and close to $T_{amb}$.

**Table 1**
Main dimensions and properties of TCAC analyzed.

| Parameter | Cable 1 | Cable 2 | Parameter | Cable 1 | Cable 2 |
|---|---|---|---|---|---|
| $V_n$ (kV) | 132 | 220 | $N$ | 114 | 110/119 |
| $I_{max}$ (A) | 732 | 655 | $L_a$ (m) | 3.5 | 3/2.3 |
| $S_n$ (mm²) | 800 | 500 | $L_c$ (m) | 2.8 | 3.5 |
| $D_c$ (mm) | 35 | 26.2 | Armor twist | contra. | contra./uni. |
| $D_s$ (mm) | 87.6 | 83.4 | $T_{amb}$ (°C) | 5 | 20 |
| $t_s$ (mm) | 3.7 | 2.9 | $\sigma_c$ (MS/m) | 51 | 59 |
| $D_{core}$ (mm) | 92.4 | 89.2 | $\sigma_s$ (MS/m) | 4.5 | 4.5 |
| $D_a$ (mm) | 214.6 | 211/228 | $\sigma_a$ (MS/m) | 5.2 | 4.03 |
| $d_a$ (mm) | 5.6 | 5.6/5.6 | $\mu_r$ | Fig. 5 | 300 |

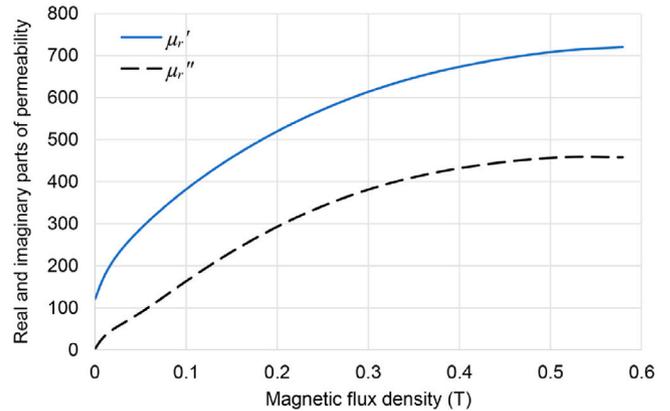

**Fig. 5.** Complex relative permeability for a LG steel.

It should be noted that Cable 1 has a single steel-layered armor (StS) in contralay configuration, while Cable 2 is doubled armored (StD), being the inner armor in contralay and the outer one in unilay (both having different values of $N$ and $L_a$). Nonetheless, the number of case studies is further extended by considering a single-layered armor also for Cable 2, assuming all the wires made of steel (StS) or combined with PE separators (St+PE) (Fig. 6). The case of a non-magnetic armor ($\mu_r = 1$, StA) is also considered, although it is equivalent to an unarmored cable, as will be discussed later.

## 4. Experimental validation of the USM

During the tests developed in [31], the authors of this work had the opportunity of taking MF measurements on Cable 1 that are here employed for the experimental validation of the USM. As shown in Fig. 7, the cable was suspended at 1.24 m above the ground. Measurements were taken for 50 Hz and 120 Hz at different heights and distances from the cable axis (measurement lines ML1, ML2, ML3 and ML4) with a 3-axis EMDEX II MF meter, having a resolution of 0.01 $\mu$T in the range from 0.01 to 300 $\mu$T at power frequency.

Thus, for 50 Hz, a phase current of 745 A and sheaths in SP, all measured and computed MF values at every point in ML1 to ML4 are represented together in Fig. 8 as a function of the distance to the cable axis (logarithmic scales employed in MF and horizontal axes for better visualization). The relative differences between measurements and simulations are also provided in the secondary axis. Similarly, Fig. 9 represents the results for both 50 Hz and 120 Hz when sheaths are in SB (phase currents of 745 A and 304 A, respectively).

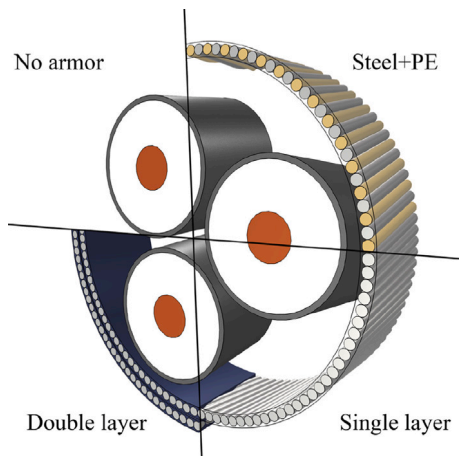

**Fig. 6.** Different configurations for the armor.

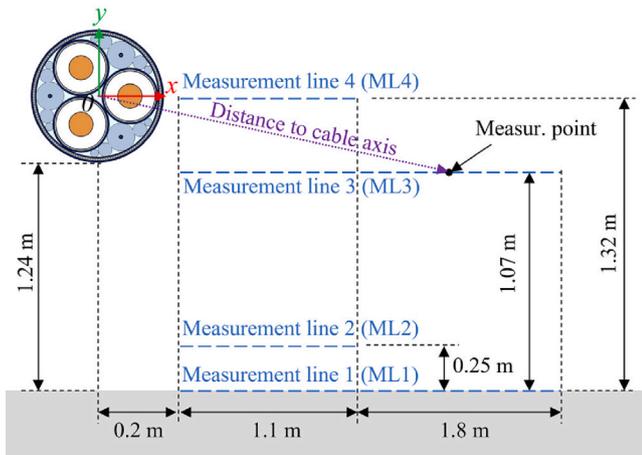

**Fig. 7.** MF measurement lines employed during Cable 1 test.

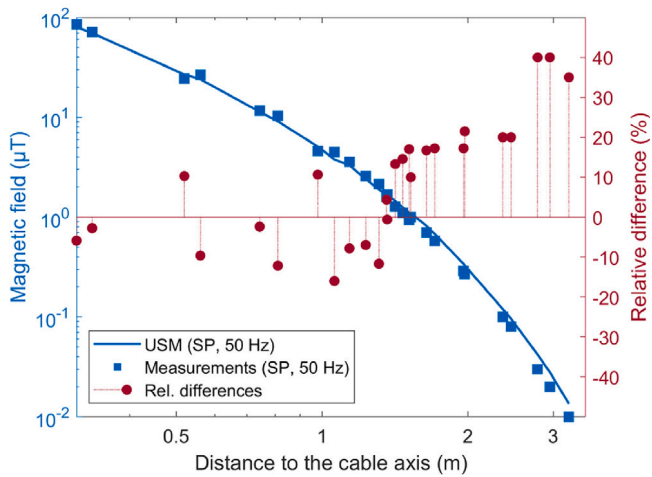

**Fig. 8.** SP (50 Hz, 745 A): Measured and computed MF values at different distances from Cable 1 axis, including its relative difference.

As can be seen, besides all possible uncertainties inherent to experimental setups (deviations in material and geometrical parameters,

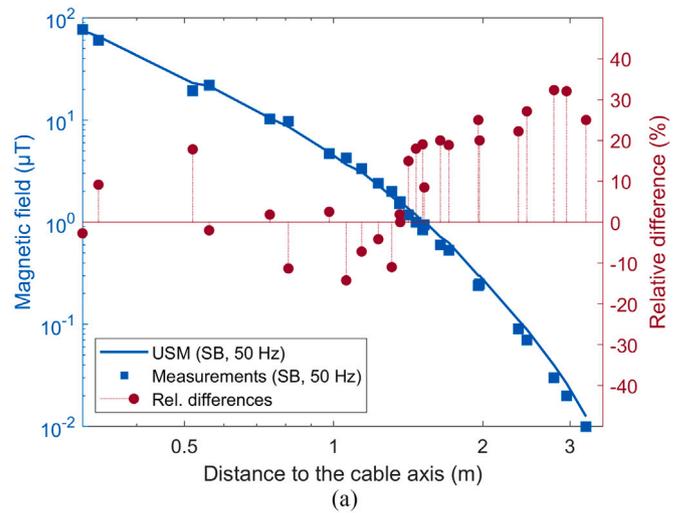

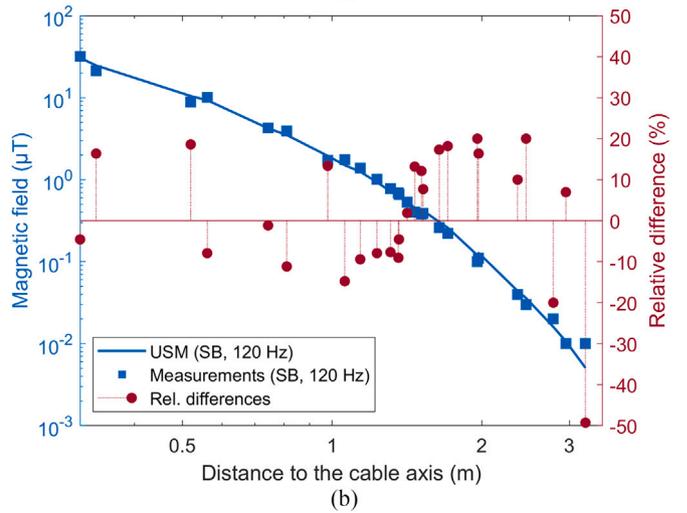

**Fig. 9.** SB: Measured and computed MF values at different distances from Cable 1 axis for (a) 50 Hz (745 A) and (b) 120 Hz (304 A), including its relative difference.

**Table 2**
Cable 1 (SB): Measured, calculated and relative difference in the sheath current values.

|  | 50 Hz (745 A) | 120 Hz (304 A) |
|---|---|---|
| Measured (A) | 187 | 136 |
| USM (A) | 186.3 | 138.9 |
| Difference (%) | −0.37 | 2.13 |

sampling errors, etc.), there is a reasonably good match between measurements and simulation results, with relative differences being below 20% in most of the cases, although it increases with the distance. These differences are a consequence of two main factors. On one hand, due to the resolution of the MF meter, which is in the same order of the MF levels to be measured at far distance from the TCAC (0.01 $\mu$T). On the other hand, due to a small imbalance observed in the phase currents ($I_s$) during the experimental tests, as reported in [31]. This disturbs the MF distribution in both the SP and SB cases, giving rise in the latter case to a small imbalance in the sheath current, as well as a net current (below 3 A) flowing through the armor. In any case, relative differences below 3% were observed in the sheath current, as summarized in Table 2.

Having all this in mind, it can be concluded that the USM is accurate enough for the analysis of the MF distribution caused by TCACs.

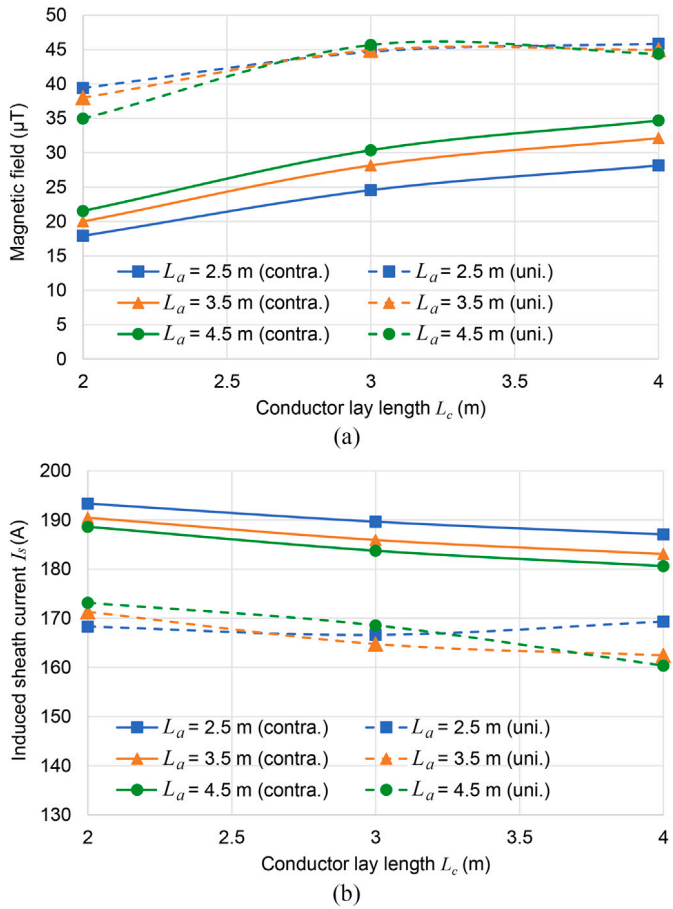

**Fig. 10.** Cable 1: Impact of $L_a$ and $L_c$ on (a) the MF at 0.5 from the cable and (b) the induced sheath current.

## 5. Influence of cable design on the MF

Through the USM, an in-depth parametric analysis is developed next for Cable 1 to show how the cable design influences the MF emissions at 0.5 m from the TCACs axis (all parameters remain as in Table 1 unless they are selected for the analysis).

### 5.1. Conductor and armor twisting

The impact of $L_a$ and $L_c$ on the MF levels at 0.5 m from Cable 1 axis are shown in Fig. 10a. It can be seen that lower MF emissions are obtained by increasing the relative twisting between the armor wires and the phase conductors (shorter values for $L_a$ and $L_c$), especially due to $L_c$. Conversely, the impact of $L_a$ depends on the twisting direction, having almost no influence in the unilay case. In any case, noticeably lower MF values are derived for a cable in contralay configuration (about half of those of the unilay case), since this encourages both the flux shunting into the armor wires and the induced sheath currents (Fig. 10b). Due to these results, in the following only the contralay configuration is considered for the analysis.

However, it is important to notice that, although decreasing $L_a$ and $L_c$ results in lower MF emissions, this also leads to higher power losses in the armor and sheaths, and hence higher cable resistance ($R$), as is shown in Fig. 11 (vertical axes not to scale for better visualization), where a similar impact is observed also on the cable inductive reactance ($X$). Therefore, cable designers must observe a balance between these opposite effects.

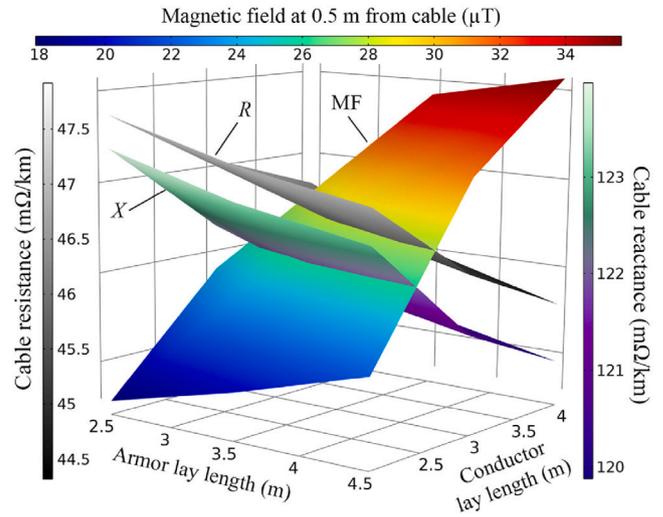

**Fig. 11.** Cable 1: Impact of $L_a$ and $L_c$ on the values of $R$, $X$ and the MF at 0.5 m.

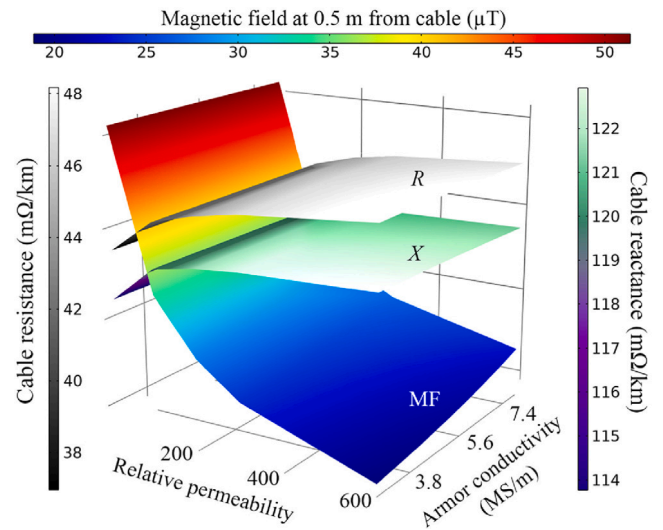

**Fig. 12.** Cable 1: Impact of $\sigma_a$ and $\mu_r$ on the values of $R$, $X$ and the MF at 0.5 m.

### 5.2. Material properties of the armor wire

The properties of the armor wires have a key role when minimizing MF emissions in TCACs. Thus, Fig. 12 shows how a great reduction is obtained in the MF values at 0.5 m from Cable 1 as $\mu_r$ increases, as expected, since this parameter enhances the flux shunting mechanism. Conversely, the effect of $\sigma_a$ is only noticeable when $\mu_r \geq 300$. Nonetheless, as in the preceding analysis, this also results in higher values for $R$ (losses) and $X$, so an appropriate optimization of the armor properties would be required.

### 5.3. Sheath and armor dimensions

The impact of sheaths and armor dimensions on the MF levels at 0.5 m from Cable 1 is represented in Fig. 13a, where different values are considered for $t_s$, $d_a$ and $N$ ($N$ varies accordingly with $d_a$ to keep the distance between the wires constant). As can be seen, the MF caused by Cable 1 can be reduced by increasing the total cross-section in sheaths and armor (higher values for $t_s$ and $d_a$), since this enhances the flux shunting effect while increasing $I_s$ also. However, this also leads to higher values in $R$ (losses), although $X$ tends to decrease with these parameters (Fig. 13b).

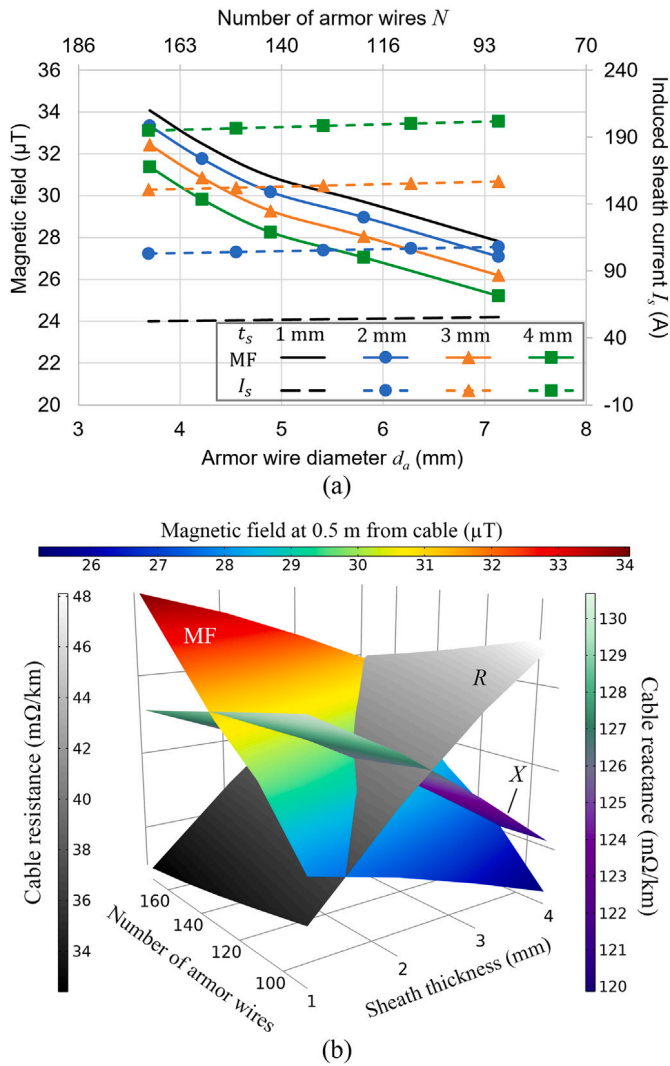

**Fig. 13.** Cable 1 ($\mu_r = 300$): Impact of $N$ ($d_a$) and $t_s$ on (a) $I_s$ and the MF at 0.5 m from the cable; (b) the MF, $R$ and $X$.

**Table 3**
Influence of material and geometrical parameters on MF emissions, $R$, $X$, and $I_s$ (contralay).

|  | MF | $R$ | $X$ | $I_s$ |
|---|---|---|---|---|
| ↓ $L_a$ | ↓ | ↑ | ↑ | ↑ |
| ↓ $L_c$ | ↓ | ⇓ | ↑ | ↑ |
| ↓ $\sigma_a$ | ↓ | ≈ | ≈ | ≈ |
| ↑ $\mu_r$ | ↓ | ⇓ | ↑ | ↑ |
| ↑ $t_s$ | ↓ | ⇑ | ↓ | ⇑ |
| ↑ $d_a$ (↓ $N$) | ↑ | ↑ | ↑ | ↑ |

As a conclusion, Table 3 summarizes, for the contralay configuration, the impact of the main material and geometrical parameters on reducing the MF emissions in TCACs, as well as side effects on $R$, $X$ and $I_s$.

### 5.4. Armor layout

The armor layout strongly influences the MF emitted by TCACs, as shown in Fig. 14, where it is represented the MF distribution obtained when different types of armor layouts are employed in Cable 2 (the most external armor is removed for the non-double-layered cases, and half of the armor wires are removed for the St+PE case). The performance, in terms of MF reduction, provided by each armor layout can be quantified through the shielding factor, defined as

$$SF = \frac{B_0}{B}, \tag{5}$$

where the MF distribution caused by the unarmored cable ($B_0$) is taken as the reference to evaluate the resulting MF ($B$) obtained through the other configurations (StA, St+PE, StS and StD).

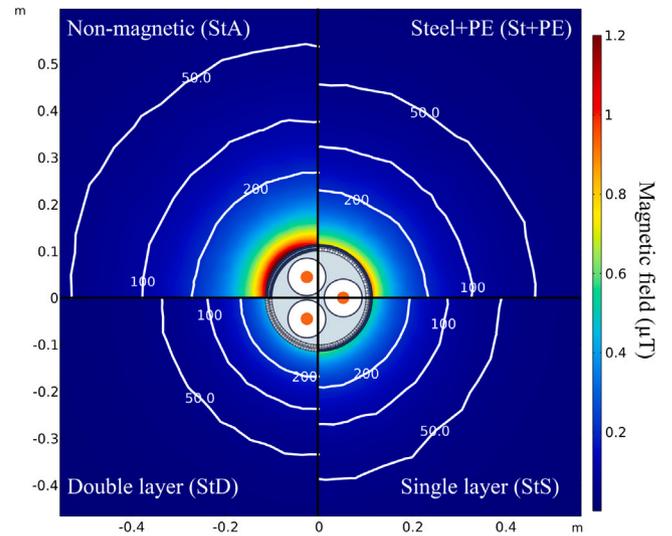

**Fig. 14.** MF distribution obtained for different armor layouts in Cable 2.

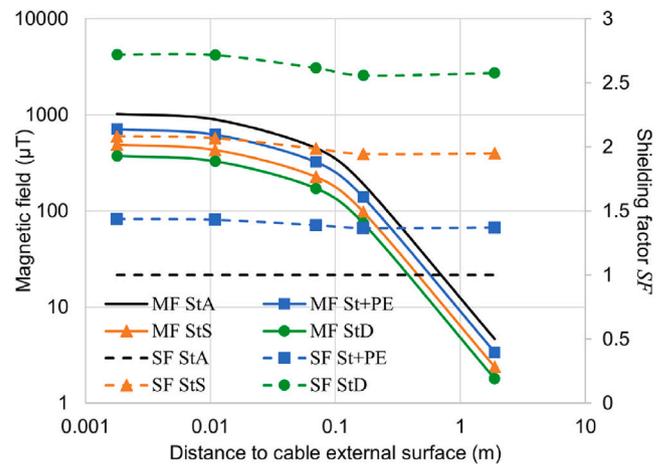

**Fig. 15.** Cable 2: MF and SF evolution with the distance for different armor layouts.

Results are represented in Fig. 15, where it can be seen a $SF = 1$ for the StA armor ($\mu_r = 1$), meaning that the MF levels provided by this case and the unarmored configuration are virtually the same, even though the former has a conductive steel armor. This is because in the non-magnetic armor there is no flux shunting effect and, moreover, eddy currents are negligible. Conversely, the St+PE configuration reduces the MF in 1.4 times, and the StS case in 2 times, although the greatest reduction is observed for the StD case with $SF = 2.7$.

Therefore, as expected, the StD case is the best choice in terms of reducing MF emissions in TCACs due to the higher number of ferromagnetic steel wires. In contrast, as commented earlier, a greater value in $N$ increases the cable electrical parameters and the induced sheath current from the StA to the StD cases, as observed in Table 4.

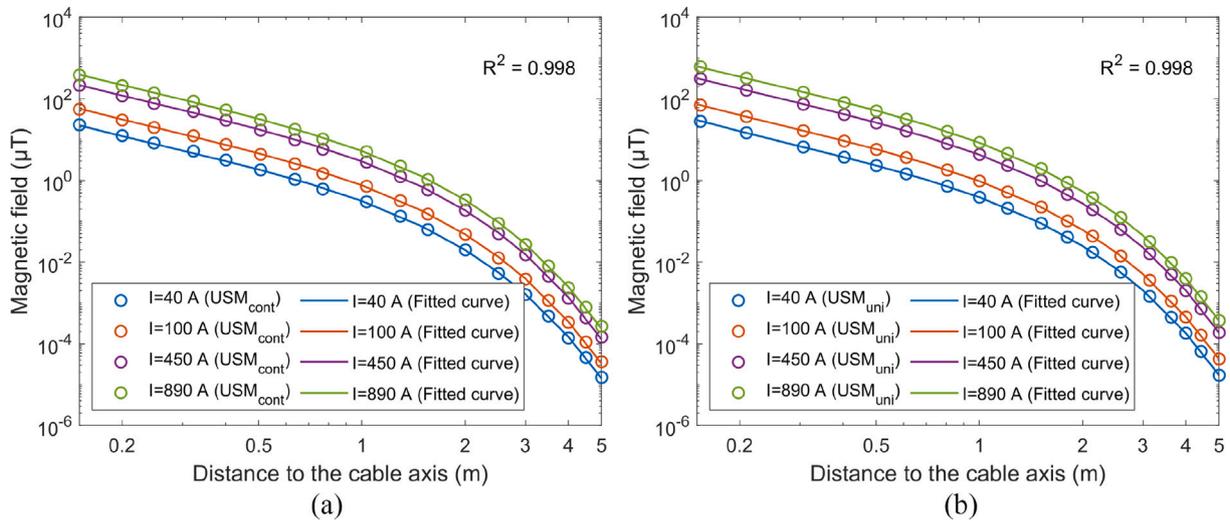

**Fig. 16.** MF evolution with the distance for different phase currents derived from the USM and the fitted curves for Cable 1 in (a) contralay and (b) unilay configurations.

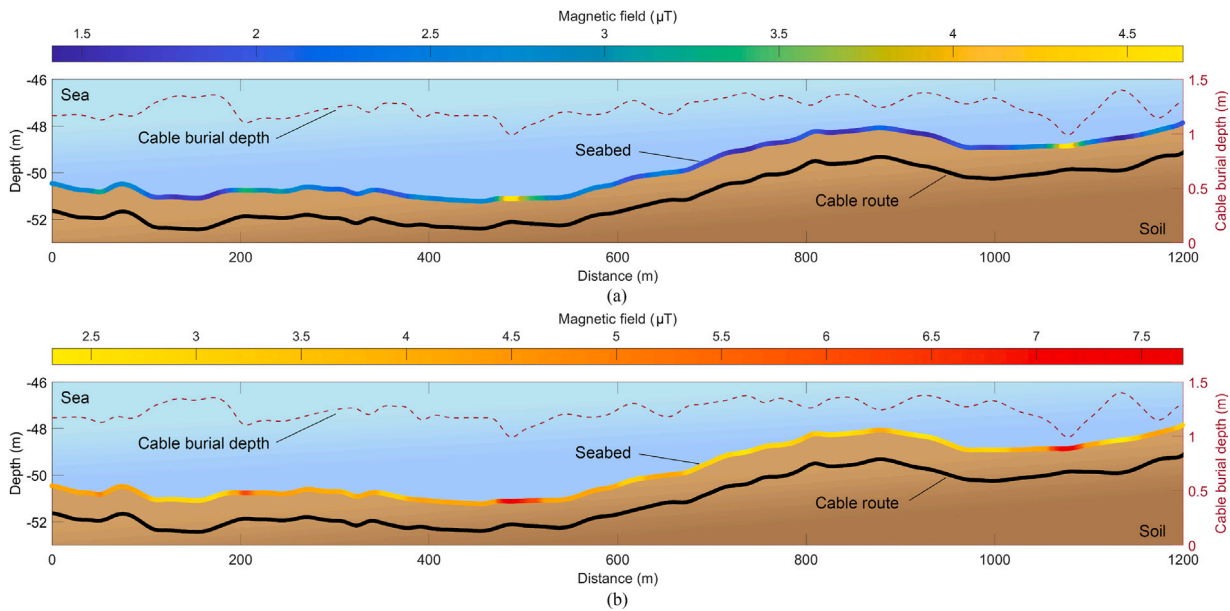

**Fig. 17.** Estimated MF emissions at the seabed surface along the Cable 1 route for (a) contralay and (b) unilay configuration (745 A of phase current).

**Table 4**
Cable 2 (SB): Sheath current, series resistance and reactance for different armor layouts.

|  | StA | St+PE | StS | StD |
|---|---|---|---|---|
| Sheath current (A) | 82.2 | 97.3 | 107.2 | 109.4 |
| Resistance (mΩ/km) | 53.3 | 60.63 | 63.18 | 64.45 |
| Reactance (mΩ/km) | 133.2 | 138.6 | 143.7 | 144.6 |

## 6. Evaluation of MF emissions along the cable route

The benefits of the USM can be easily employed for the evaluation of the environmental impact, in terms of MF emissions, of a particular TCAC along the whole cable route. To this aim, a parametric study can be developed through the USM for obtaining different MF-distance profiles for a range of phase currents (Fig. 16). From all these data, an approximate expression can be derived ($B_{fit}$), through a curve-fitting technique, to evaluate the MF at any distance from the TCAC ($r$) as a function of the phase current, in the form of

$$B_{fit}(\mu T) = 10^{[k_1 \cdot \exp(k_2 \cdot \log_{10}(r)) + k_3 \cdot \exp(k_4 \cdot \log_{10}(r))]}, \quad (6)$$

being

$$k_i = \alpha_{1i} \cdot I^{\alpha_{2i}} + \alpha_{3i}, \quad i = 1, 2, 3, 4. \quad (7)$$

The values for coefficients $\alpha_{1i}$, $\alpha_{2i}$ and $\alpha_{3i}$ are illustrated in Table 5 for Cable 1 in contralay and unilay configurations. The results derived from the USM ($B_{USM}$) and the proposed expression for different phase currents and armor twisting have a very good match (Fig. 16, logarithmic scales are employed for clarity), where the corresponding coefficient of determination, $R^2$, is included to show the goodness of the adjustments. Some of these results are also summarized in Table 6 (for the contralay case) to show how the relative differences ($\epsilon$) between $B_{USM}$ and $B_{fit}$ are well below 10% in all the cases.

Using this procedure, the MF emissions in the subsea environment can be obtained for different situations, such as cable loading, burial depth, cable design, etc. Fig. 17 shows an example of application for Cable 1, representing the MF levels at the seabed surface for both, contralay (Fig. 17a) and unilay (Fig. 17b) configurations, including variations in the cable burial depth. As expected, higher MF values are observed for the unilay case, that almost double those of the contralay configuration. On the other hand, hot spots are also highlighted in both cases, indicating those sections where the cable is closer to the seabed surface (shorter burial depth). Therefore, this tool helps not only in evaluating the environmental impact in terms of MF emissions, but also in developing new techniques for detecting/tracking buried cables, detecting exposed sections or burial depth estimation.

**Table 5**
Coefficients for evaluating MF emissions in Cable 1 for contralay and unilay configurations.

| | Contralay | | | | Unilay | | | |
|---|---|---|---|---|---|---|---|---|
| | $k_1$ | $k_2$ | $k_3$ | $k_4$ | $k_1$ | $k_2$ | $k_3$ | $k_4$ |
| $\alpha_{1i}$ | −1.376 | −1.275 | 12.4 | −2.998 | −1.08 | 3.817 | 4.885 | −3.542 |
| $\alpha_{2i}$ | −0.4061 | −0.4346 | 0.02313 | −0.3657 | −0.3319 | 0.01328 | 0.05577 | −0.4098 |
| $\alpha_{3i}$ | −0.7711 | 2.479 | −12.93 | −0.4041 | −0.7132 | −1.723 | −5.383 | −0.3957 |

**Table 6**
Cable 1 (contralay): Computed and fitted MF values, and their relative differences, at different distances from the cable axis and phase currents (superscripts in Amperes).

| $r$ (m) | 0.15 | 0.5 | 1 | 1.5 | 2 | 2.5 | 3 | 4 | 5 |
|---|---|---|---|---|---|---|---|---|---|
| $B_{USM}^{40}$ (µT) | 23.08 | 1.81 | 0.30 | 0.076 | $1.99 \cdot 10^{-2}$ | $5.26 \cdot 10^{-3}$ | $1.60 \cdot 10^{-3}$ | $1.39 \cdot 10^{-4}$ | $1.48 \cdot 10^{-5}$ |
| $B_{fit}^{40}$ (µT) | 22.31 | 1.85 | 0.31 | 0.074 | $1.98 \cdot 10^{-2}$ | $5.58 \cdot 10^{-3}$ | $1.63 \cdot 10^{-3}$ | $1.48 \cdot 10^{-4}$ | $1.42 \cdot 10^{-5}$ |
| $\varepsilon^{40}$ (%) | −3.3 | 2.4 | 4.5 | −1.6 | −0.4 | 6.0 | 1.9 | 6.7 | −4.2 |
| $B_{USM}^{100}$ (µT) | 55.05 | 4.34 | 0.72 | 0.181 | 0.048 | $1.26 \cdot 10^{-2}$ | $3.84 \cdot 10^{-3}$ | $3.33 \cdot 10^{-4}$ | $3.44 \cdot 10^{-5}$ |
| $B_{fit}^{100}$ (µT) | 57.39 | 4.60 | 0.76 | 0.179 | 0.048 | $1.34 \cdot 10^{-2}$ | $3.90 \cdot 10^{-3}$ | $3.48 \cdot 10^{-4}$ | $3.25 \cdot 10^{-5}$ |
| $\varepsilon^{100}$ (%) | 4.3 | 5.9 | 5.7 | −1.0 | −0.1 | 6.2 | 1.6 | 4.6 | −5.6 |
| $B_{USM}^{450}$ (µT) | 212.88 | 16.95 | 2.81 | 0.708 | 0.186 | $4.93 \cdot 10^{-2}$ | $1.50 \cdot 10^{-2}$ | $1.30 \cdot 10^{-3}$ | $1.36 \cdot 10^{-4}$ |
| $B_{fit}^{450}$ (µT) | 215.55 | 17.79 | 2.92 | 0.692 | 0.185 | $5.24 \cdot 10^{-2}$ | $1.53 \cdot 10^{-2}$ | $1.38 \cdot 10^{-3}$ | $1.28 \cdot 10^{-4}$ |
| $\varepsilon^{450}$ (%) | 1.3 | 4.9 | 4.0 | −2.2 | −0.6 | 6.3 | 2.2 | 5.5 | −6.0 |
| $B_{USM}^{890}$ (µT) | 379.61 | 30.40 | 5.04 | 1.27 | 0.334 | $8.85 \cdot 10^{-2}$ | $2.69 \cdot 10^{-2}$ | $2.34 \cdot 10^{-3}$ | $2.47 \cdot 10^{-4}$ |
| $B_{fit}^{890}$ (µT) | 388.70 | 32.18 | 5.26 | 1.24 | 0.333 | $9.45 \cdot 10^{-2}$ | $2.77 \cdot 10^{-2}$ | $2.49 \cdot 10^{-3}$ | $2.32 \cdot 10^{-4}$ |
| $\varepsilon^{890}$ (%) | 2.4 | 5.9 | 4.3 | −2.1 | −0.4 | 6.8 | 2.9 | 6.5 | −6.2 |

## 7. Conclusions

This paper makes an in-depth analysis of the MF generated by TCACs, including the effects of cable design on the resulting MF emissions. To this aim, 3D-FEM simulations are performed through the USM proposed in previous studies, since 2D-FEM simulations lead to inaccurate results derived from the omission of the armor and conductor twisting. The accuracy of the USM is first evaluated through MF experimental measurements taken for an actual 132 kV, 800 mm$^2$ TCAC, resulting in relative differences typically below 20% despite typical uncertainties in measurements and input data.

Once validated, the USM is employed for analyzing the impact of the main geometrical and material parameters on the MF values at 0.5 m from the cable. Results show how the MF levels reduce with the power core twisting (shorter lay length) and the armor permeability. Both parameters also influence the induced sheath currents, that together with the sheath thickness increase the mitigation effect and, hence, reduce the MF levels.

Regarding the armor, it is observed that contralay configuration generates less MF than unilay, being this aspect much more influencing than the lay length of the armor. The number and diameter of armor wires also help in reducing the MF emissions. In this sense, it is also observed that the armor layout strongly influences the results. Thus, a TCAC where non-magnetic steel wires are employed in the armor behaves like an unarmored cable, while the MF can be reduced in 1.4 times (compared with the unarmored cable) if a single-layer armor combines magnetic steel wires with PE separators. However, the best results are obtained when fully steel-wired armors are employed, especially in the case of double-layered armors, leading to MF values 2.7 times lower than in the unarmored case.

Finally, through an application example, it is shown how the USM is a valuable tool for evaluating the environmental impact caused by TCACs in terms of MF emissions, providing remarkable information for obtaining a proper cable design. Moreover, this tool also serves as a platform for the development of new techniques devoted to cable tracking, detection of exposed sections or burial depth estimation.

## CRediT authorship contribution statement

**Juan Carlos del-Pino-López:** Conceptualization, Methodology, Validation, Formal analysis, Investigation, Data curation, Writing – original draft. **Pedro Cruz-Romero:** Conceptualization, Formal analysis, Data curation, Writing – review & editing. **Juan Carlos Bravo-Rodríguez:** Investigation, Data curation, Writing – review & editing.

## Declaration of competing interest

The authors declare that they have no known competing financial interests or personal relationships that could have appeared to influence the work reported in this paper.

## Data availability

Data will be made available on request.

## Acknowledgments

This research is part of the project ENE2017-89669-R, funded by MCIN/AEI/10.13039/501100011033, by ERDF "A way of making Europe" and by the Universidad de Sevilla (VI PPIT-US) under grant 2018/00000740.

The authors would like to acknowledge and thank Jarle Bremnes and Marius Hatlo for bringing the opportunity to take MF measurements during their tests in the Halden factory.